\documentclass[preprint,authoryear,12pt]{elsarticle}

\usepackage{graphicx}
\usepackage{amsmath}
\usepackage{amssymb}
\usepackage{url}
\providecommand{\degr}{\ensuremath{^{\circ}}}
\providecommand{\upi}{\pi}

\journal{Icarus}

\begin{document}

\begin{frontmatter}

\title{The rotation period of the near-Earth comet (875163)\,1998\,SH$_2$ and
the limits of a single-apparition phase curve}

\author[iac,ull]{Javier Licandro\corref{cor1}}
\ead{jlicandr@iac.es}
\author[lb,iac,ull]{Miguel R. Alarcon}
\author[lb,iac,ull]{Miquel Serra-Ricart}
\cortext[cor1]{Corresponding author.}

\address[iac]{Instituto de Astrof\'isica de Canarias, V\'ia L\'actea s/n,
  E-38205 La Laguna, Tenerife, Spain}
\address[ull]{Departamento de Astrof\'{\i}sica, Universidad de La Laguna,
  38206 La Laguna, Tenerife, Spain}
\address[lb]{Light Bridges, Observatorio Astron\'omico del Teide, Carretera del
  Observatorio del Teide s/n, G\"u\'imar, Canarias, Spain}

\begin{abstract}
The near-Earth object (875163)\,1998\,SH$_2$ was recently shown to be a weakly
active comet: its 1998--2025 astrometry
reveals a non-gravitational acceleration too large to be radiative in origin,
and large-aperture imaging during its 2025 apparition revealed a faint tail
\citep{farnocchia2026}. No rotation period had been reported. We derive one from
dense TTT2 photometry obtained on five nights of the same apparition: from 2392
calibrated measurements we find a synodic rotation period
$P = 6.970 \pm 0.020$\,h. The folded light curve is markedly bimodal
($A_2/A_1 \simeq 6.4$) with a peak-to-peak amplitude of $\simeq 0.18$\,mag,
implying $a/b \gtrsim 1.18$. Combined with the published radar Doppler
bandwidth, the period is consistent with an elongated body whose rotation the
radar alone could not constrain. Over the narrow
observed phase-angle range ($3.7\degr$--$9.3\degr$) the IAU $H,G$ and
$H,G_1,G_2$ slope parameters are not physically constrained. The inter-night
magnitude gradient, $\beta_{\rm app} = 0.074 \pm 0.005$\,mag\,deg$^{-1}$, is steeper than the
default $G=0.15$ and lies in the upper part of the Jupiter-family comet (JFC)
distribution (median $0.046$; \citealt{kokotanekova2017}); because the object
was weakly and increasingly active during the run, we treat this value as an
\emph{upper limit} to the intrinsic surface phase slope. Under the population
prior of little or no opposition surge, as found in available JFC nucleus
phase curves, the phase behaviour is best described by a linear law, giving a provisional
$H_{\rm Lum}(1,1,0) \approx 20.9$--$21.0$ over $3.7\degr$--$9.3\degr$ in the Luminance band.
\end{abstract}


\end{frontmatter}

\section{Introduction}
The classical distinction between asteroids and comets has blurred into a
continuum of hybrid objects \citep{jewitt2024,seligman2024}. Among these are the
so-called `dark comets': near-Earth objects that show no obvious coma yet
exhibit non-gravitational accelerations inconsistent with radiation forces,
suggesting weak, undetected outgassing. Near-Earth object (875163)\,1998\,SH$_2$
is a defining example. \citet{farnocchia2026} showed that its long (1998--2025)
astrometric arc requires a transverse non-gravitational acceleration,
$A_2 = (-1.4 \pm 0.1)\times 10^{-11}$\,m\,s$^{-2}$, an order of magnitude larger
than the maximum plausible Yarkovsky effect for its size, and subsequently
detected a faint dust tail with the CFHT, VLT and Danish 1.54\,m telescopes. The
physical properties of 1998\,SH$_2$  reinforce the cometary interpretation: a NEOWISE diameter of
$380 \pm 57$\,m and a very low albedo of $0.058 \pm 0.024$ \citep{masiero2017}, and a Tisserand
parameter with respect to Jupiter of $T_J = 2.9$, typical of Jupiter-family
comets (JFCs).

In dynamical terms 1998\,SH$_2$ was considered as an asteroid in a cometary orbit (ACO).
Following the classification of \citet{tancredi2014}, ACOs occupy unstable,
Jupiter-family- or Halley-type orbits and are the strongest dynamical candidates
for dormant or extinct comets. Their surfaces have been characterized in a
series of studies that consistently point to a cometary nature. Spectroscopically
they are dominated by primitive, featureless red material: essentially all
Damocloids and $\sim\!60$ per cent of JFC-ACOs are D-types, the remainder mostly
X-types, with spectral slopes and an absence of hydration features matching
cometary nuclei \citep{licandro2008,licandro2018}. Thermally, their
WISE-derived geometric albedos are low and narrowly distributed,
$\bar{p}_V = 0.05 \pm 0.02$, indistinguishable from the $\sim\!0.04$ of JFC
nuclei \citep{licandro2016}. The NEOWISE albedo of 1998\,SH$_2$
($0.058 \pm 0.024$) places it squarely within this population, so that (even
before the activity detection) it was a textbook dormant-comet candidate; the
tail found by \citet{farnocchia2026} now confirms that inference for this
individual object. This dark, D-type-like surface also underpins the phase-curve
prior we adopt in Section~\ref{sec:phase}.

Several features make 1998\,SH$_2$ a particularly valuable target. It is one of
the very few NEOs for which cometary activity has been not merely
suspected but confirmed \citep{farnocchia2026}, with direct implications for the impact-hazard assessment of objects that outgas. It is also among the smallest and
closest of these objects: its 2025 close approach (0.02\,au) placed it within reach of
both Goldstone radar and metre-class photometry during the same weeks that its
activity was detected. Yet no rotation period has been reported, leaving
its spin state, shape and phase behaviour unconstrained. This combination is rare:
a confirmed, dynamically comet-like body, bright enough for dense time-series
photometry, observed contemporaneously with its incipient activity and with an
independent radar size and thermal-IR albedo already in hand. It therefore offers
an unusually complete opportunity to tie together rotation, shape and surface
properties for an object caught in the act of transitioning between the
asteroidal and cometary regimes, and, more generally, to test what a single,
narrow-phase-angle apparition can reveal about such weakly active bodies.

Rotational light curves of such objects constrain their spin and shape and, in
favourable geometries, support radar shape modelling. For 1998\,SH$_2$ the
Goldstone radar detection of 2025 September 2 yielded a Doppler bandwidth of
7\,Hz but, without an independent period, could only bound the rotation to
$P < 5.4$\,h under the assumption of a spherical, non-elongated 380\,m body; no
rotation period has been reported \citep{farnocchia2026}. In this paper we determine that
period from photometry obtained during the same apparition, and examine what a
single, narrow-phase-angle apparition can and cannot say about the object's
magnitude--phase relation -- a question sharpened by the object's dark, cometary
nature.

This paper is organized as follows: Section~\ref{sec:obs} describes the observations and the observational circumstances; Section~\ref{sec:analysis} presents the photometric results and the rotational and phase-curve properties derived from the time-series photometry; Section~\ref{sec:results} discusses the results in the context of weakly active comets; and Section~\ref{sec:conclusions} summarizes our conclusions.

\section{Observations}
\label{sec:obs}
Photometry was acquired on 2025 September 9, 11, 12, 13 and 14 with TTT2, one of the two 0.80-m $f$/6.85 units of the Two-meter Twin Telescope (TTT\footnote{\url{https://ttt.iac.es/}}) at the Teide Observatory, Canary Islands, Spain ($28^\circ18'04''$~N, $16^\circ30'38''$~W, alt.\ 2362~m). The facility, operated by Light Bridges S.L., comprises two identical 2.0-m $f$/6 Ritchey-Chr\'etien telescopes (TTT3/4) and two 0.80-m replicas (TTT1/2), all robotic, on Alt-Az mounts and with their instruments at the Nasmyth foci; TTT2 has been operational since 2023. The observations used the FERVOR-M imager (Fast Embedded-sCMOS Robotic Visible Observatory for Rapid transients), based on a back-illuminated Sony IMX455 sCMOS sensor \citep[$3.76\,\mu$m pixels, readout noise $1.7\,e^-$, negligible dark current and essentially no readout overhead;][]{Alarcon2023}, which delivers a field of view of $22.6'\times 15.0'$ at a plate scale of $0.283''~\mathrm{px}^{-1}$, operated with $2\times2$ binning. All frames were taken through a broadband UV-IR-cut (``Luminance'') filter with a flat effective transmission between 400 and 700~nm, approximately equivalent to the combined Sloan $g'+r'$ bandpass \citep{ATLASTeide}.

The images were bias- and flat-field corrected following standard procedures, and then analyzed with \textsc{GPUPhot}, the GPU-native point-source detection and photometry pipeline of the TTT  \citep{Alarcon2026GPUPhot, LemesPerera2026GPUPhot}, which performs source detection with a spatially variable, empirically derived model of the point spread function (PSF). The flux of each source is measured in a circular aperture whose radius is set automatically, frame by frame, to the value that maximizes the signal-to-noise ratio for the local PSF width (bounded to the range $0.75$--$5$ times the full width at half maximum, FWHM), after which the corresponding PSF-based aperture correction is applied  \citep{Alarcon2026GPUPhot, LemesPerera2026GPUPhot}. Each frame is finally calibrated astrometrically with Astrometry.net \citep{lang2010} and photometrically against Pan-STARRS\,DR1 \citep{chambers2016}. The instrumental Luminance magnitudes are referred to the synthetic combination $m = 0.516\,g_{\rm P1} + 0.484\,r_{\rm P1}$; with coefficients obtained empirically from the least-squares fit of instrumental Luminance magnitudes to PS1 g and r photometry of field stars, as  in \cite{alarcon2026}. The zero-point is fixed at unit slope from a $\sigma$-clipped set of near-solar-color calibrators appropriate to the Solar System targets of these surveys.

The photometry of 1998\,SH$_2$ was recovered from that database a posteriori, by propagating the JPL Horizons ephemeris of the object to the mid-exposure epoch of each frame and matching the predicted sky position against the calibrated
detections of that frame within a tolerance of 2 arcsec. This yields, for every exposure, the light-time-corrected epoch, the calibrated magnitude $m$ and its uncertainty, the heliocentric ($r$) and geocentric ($\Delta$) distances and the solar phase angle $\alpha$, together with the distance- and phase-reduced magnitudes computed with the $H,G$ system \citep{bowell1989} at the default slope $G=0.15$. We retained only measurements with photometric uncertainties below 0.1\,mag and visually inspected the extracted light curve to remove residual outliers (asymmetric PSFs, trails, blends). The circumstances of the five nights are summarized in Table~\ref{tab:obs}. The individual nightly light curves are shown in \ref{app:nightly}.

\begin{table}
\centering
\caption{Observational circumstances. All data were obtained with the TTT2
telescope using the FERVOR-M imager (BSI sCMOS IMX455 sensor) through 
a broadband UV/IR Cut filter (400--700~nm).
$N$ is the number of measurements retained after the quality checks described in the text, $\Delta t$ the time span of the night, and $r$,
$\Delta$ and $\alpha$ the mean heliocentric distance, geocentric distance and
solar phase-angle range; $\bar{V}$ is the mean apparent magnitude; $\tilde{\sigma}_m$ is the median per-point photometric uncertainty of the night; and the mean FWHM of the images of the night. $\tilde{\sigma}_m$ and FWHM tracks the observing conditions.}

\label{tab:obs}
\setlength{\tabcolsep}{3.8pt}
\begin{tabular}{lrrccccccc}
\hline
Night & MJD$_{\rm start}$ & $N$ & $\Delta t$ & $r$ & $\Delta$ & $\alpha$ & $\bar{V}$ & $\tilde{\sigma}_m$ &FWHM \\
(2025) & & & (h) & (au) & (au) & (deg) & (mag) & (mag) & (arcsec)\\
\hline
Sep 9  & 60927.93 & 699 & 3.0 & 1.113 & 0.107 & 9.1--9.3 & 17.00 & 0.048 & $2.09^{+0.25}_{-0.23}$ \\
Sep 11 & 60929.91 & 517 & 7.2 & 1.134 & 0.129 & 6.3--6.7 & 17.21 & 0.033 & $1.70^{+0.32}_{-0.24}$ \\
Sep 12 & 60931.05 & 412 & 3.5 & 1.145 & 0.139 & 5.3--5.5 & 17.34 & 0.028 & $1.79^{+0.40}_{-0.21}$  \\
Sep 13 & 60932.05 & 181 & 1.5 & 1.154 & 0.149 & 4.5      & 17.49 & 0.022 & $1.24^{+0.31}_{-0.11}$ \\
Sep 14 & 60932.95 & 583 & 5.6 & 1.165 & 0.159 & 3.7--3.8 & 17.55 & 0.031 & $1.56^{+0.46}_{-0.17}$ \\

\hline
\end{tabular}
\end{table}

The data set comprises $N = 2392$ measurements spanning
MJD\,60927.93--60933.18 (5.26\,d), with a median uncertainty of 0.033\,mag. The
first night was obtained under the poorest conditions of the run, with a median
per-point uncertainty of 0.048\,mag (Table~\ref{tab:obs}) and correspondingly
larger scatter in its light curve (\ref{app:nightly}); this does not
affect the period determination or the bimodality of
the folded curve (Section~\ref{sec:period}). The
geometry changed appreciably over the run: $r = 1.112$--$1.166$\,au,
$\Delta = 0.107$--$0.161$\,au and $\alpha = 3.69\degr$--$9.26\degr$. Because
$\Delta$ grows by $\sim 50$\%, the apparent magnitude carries a
$\sim 1$\,mag geometric trend that must be removed before any rotational
analysis. This window is contemporaneous with the onset of the activity reported
by \citet{farnocchia2026} that detected a weak tail on September 13, and the
coma contribution grew from $\sim 0.06$ to $0.24$\,mag between September 13 and
30. During our (earlier) run the coma was therefore weak but not necessarily
constant, a point we return to in Section~\ref{sec:phase}.

\section{Data analysis}
\label{sec:analysis}
\subsection{Reduced magnitudes}
The distance-reduced magnitude in the Luminance filter
\begin{equation}
H_{\rm Lum}(1,\alpha) = m - 5\log_{10}(r\,\Delta),
\end{equation}
removes the distance dependence but retains the phase-angle trend. As seen before, $\alpha$ varied from $3.69$ to $9.26\degr$ during our observations, so to properly compute the rotation period this trend has to be removed.


\subsection{Rotation period}\label{sec:period}

\begin{figure}
\centering
\includegraphics[width=\textwidth,height=0.75\textheight,keepaspectratio]{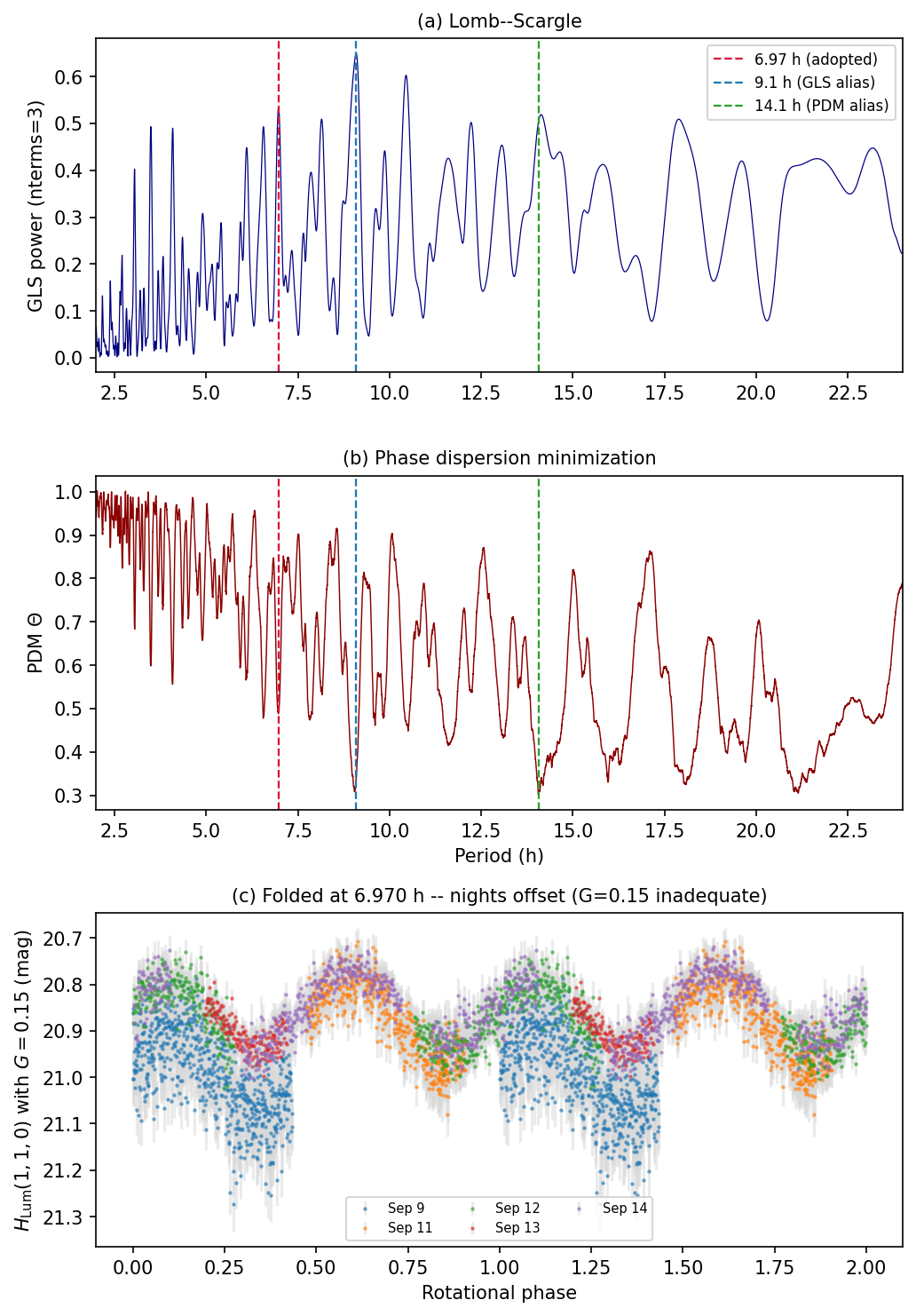}
\caption{Initial period search on the magnitudes reduced to zero phase angle with
the IAU $H,G$ function \citep{bowell1989} at the default slope $G=0.15$.
(a) Generalized Lomb--Scargle periodogram (three harmonics), whose global
maximum is the 9.1\,h daily alias; (b) phase-dispersion minimization, whose
deepest minimum is the 14.1\,h alias -- the two differ by almost exactly one
cycle per day, and the adopted period (6.97\,h, red) is present in both but is
not the strongest feature. (c) The light curve folded at 6.97\,h, coloured by
night: with the $G=0.15$ reduction the nights, taken at different phase angles,
do not share a common level (inter-night r.m.s.\ 0.05\,mag), showing that the
default slope is inappropriate for this object. The aliasing is resolved by the
continuous-night test of Fig.~\ref{fig:night}, and both the co-registration and
the periodogram become clean once the linear phase correction is applied
(Section~\ref{sec:phase}; Fig.~\ref{fig:lc}).}
\label{fig:initper}
\end{figure}

We first used the absolute magnitude reduced to zero $\alpha$ computed as
$H_{\rm Lum}(1,0) = m - 5\log_{10}(r\Delta) - \phi(\alpha)$ with the IAU $H,G$ phase
function \citep{bowell1989} and the widely used default slope $G = 0.15$.  Then we searched these
for periodicity with the generalized Lomb--Scargle (GLS) periodogram
\citep{vanderplas2018} using three Fourier harmonics, and independently with
phase-dispersion minimization (PDM; \citealt{stellingwerf1978}). Both methods are
dominated by aliasing but favour different daily aliases (Fig.~\ref{fig:initper}):
the GLS peaks at 9.1\,h and the PDM minimizes at 14.1\,h, values that differ by
almost exactly one cycle per day and are therefore mutual daily aliases of the
nightly sampling window, while the $\simeq 6.97$\,h candidate is present in both
but is the strongest in neither. Peak strength alone thus does not identify the
period; if anything, the two statistics agree only in being misled by the
window. Moreover, folding at the (adopted) 6.97\,h period already shows that the $G=0.15$ reduction does not bring
the five nights, taken at different phase angles, onto a common level
(Fig.~\ref{fig:initper}c; inter-night r.m.s.\ 0.05\,mag) -- a first indication
that the residual trend arises not from a missing phase correction but from an
inadequate phase-angle correction. A better correction is developed in Section~\ref{sec:phase}.

\begin{figure}
\centering
\includegraphics[width=\textwidth]{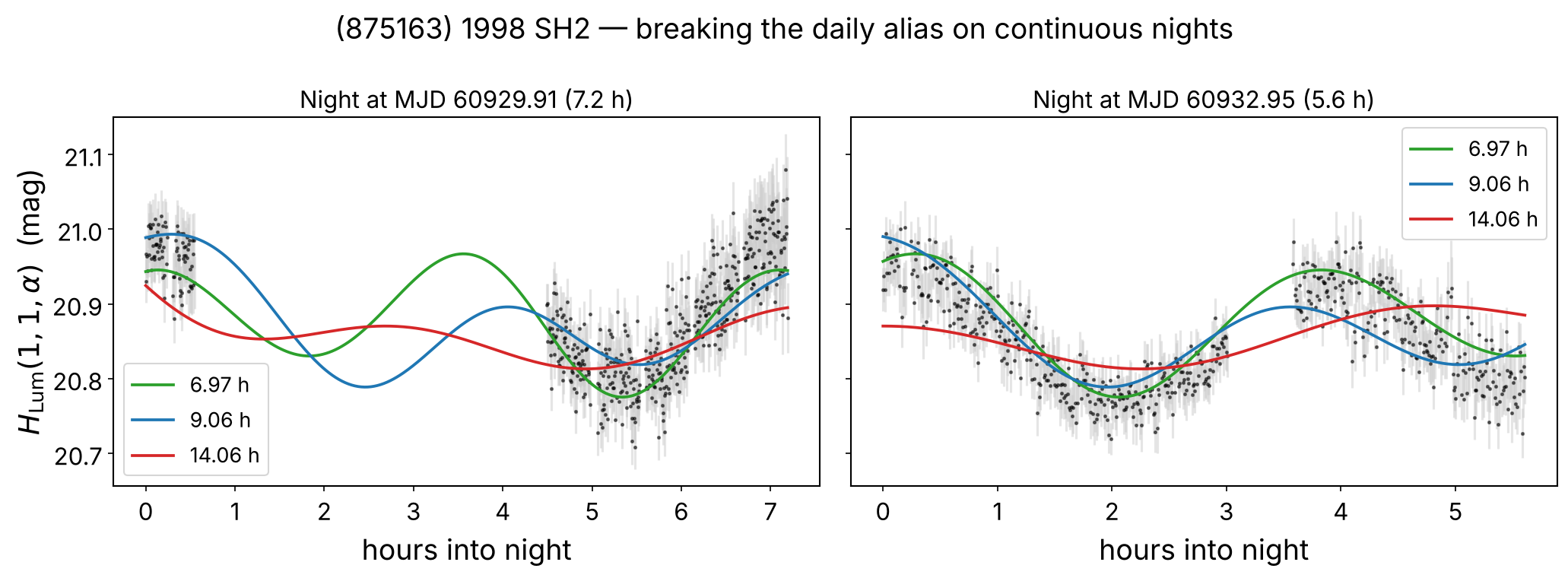}
\caption{Breaking the daily alias. Candidate-period models (a shared rotational
shape with a free per-night offset) are overplotted on the two continuous nights,
whose intra-night sampling is immune to one-cycle-per-day aliasing. Only the
6.97\,h model (green) follows the real within-night variation; the 9.06\,h and
14.06\,h aliases do not. The within-night $\chi^2/{\rm dof}$ are 0.96, 1.16 and
1.40 ($\Delta\chi^2 = 220$ and $480$; $N = 1100$).}
\label{fig:night}
\end{figure}

The degeneracy is broken by the two nights with \emph{continuous} data over 7.2\,h and
5.6\,h respectively, whose intra-night sampling is immune to daily aliasing. We fit a
three-harmonic rotation of shared shape with a free zero-point offset per night
(marginalizing the inter-night levels) and evaluate the residuals within the
two continuous nights ($N = 1100$; 8 free parameters; 1092 degrees of freedom).
Only the 6.97\,h model tracks the real within-night variation
(Fig.~\ref{fig:night}): it gives $\chi^2/{\rm dof} = 0.96$, against $1.16$ and
$1.40$ for the 9.06\,h and 14.06\,h aliases, i.e.\ $\Delta\chi^2 = 220$ and
$480$ which is decisive at this number of points. Because the per-night offsets absorb
any change in level, the test depends only on the intra-night shape. It is
therefore insensitive to the adopted phase correction: any phase law -- the
default $G = 0.15$ or any linear alternative -- leaves the within-night $\chi^2$
essentially unchanged. The selection of the 6.97\,h alias thus does not rely on
the adopted phase law. Re-fitting the three-harmonic model to the full five-night
arc at this alias refines the synodic period to $P_{\rm rot} = 6.970$\,h.

Perturbing the
magnitudes within their photometric uncertainties and re-fitting (a Monte Carlo
propagation of the periodogram peak) gives a formal $1\sigma$ uncertainty of only
$0.0011$\,h. This value is, however, unrealistically small: it measures the width
of the periodogram peak under the assumptions that the photometric errors are
correct and uncorrelated and that there is no inter-night cycle-count ambiguity,
none of which holds strictly here (the fits have $\chi^2/N \approx 2$, indicating
underestimated or correlated errors, and the daily aliasing is precisely such an
ambiguity). We therefore adopt the more conservative estimate given by the period
change that produces a visible drift of the composite light curve over the
observing arc, following the Fourier-analysis approach of \citet{harris1989},
$\sigma_P \simeq P^2\,\delta\phi/(2\,\Delta T)$, where $\Delta T \simeq 5.26$\,d is
the total arc and $\delta\phi \approx 0.1$ the largest phase drift still consistent
with a coherent folded curve. This gives $\sigma_P \approx 0.02$\,h, an order of
magnitude larger than the formal value, and we quote
\begin{equation}
P_{\rm rot} = 6.970 \pm 0.020~\mathrm{h}.
\end{equation}

Notice that we quote a synodic period throughout because the spin pole is unknown, so no sidereal period is attempted. But, in any case, over our five-night arc the synodic--sidereal difference is a factor $\sim\!8$ below our $\pm0.020$\,h uncertainty
and hence negligible.

\begin{figure}
\centering
\includegraphics[width=\textwidth]{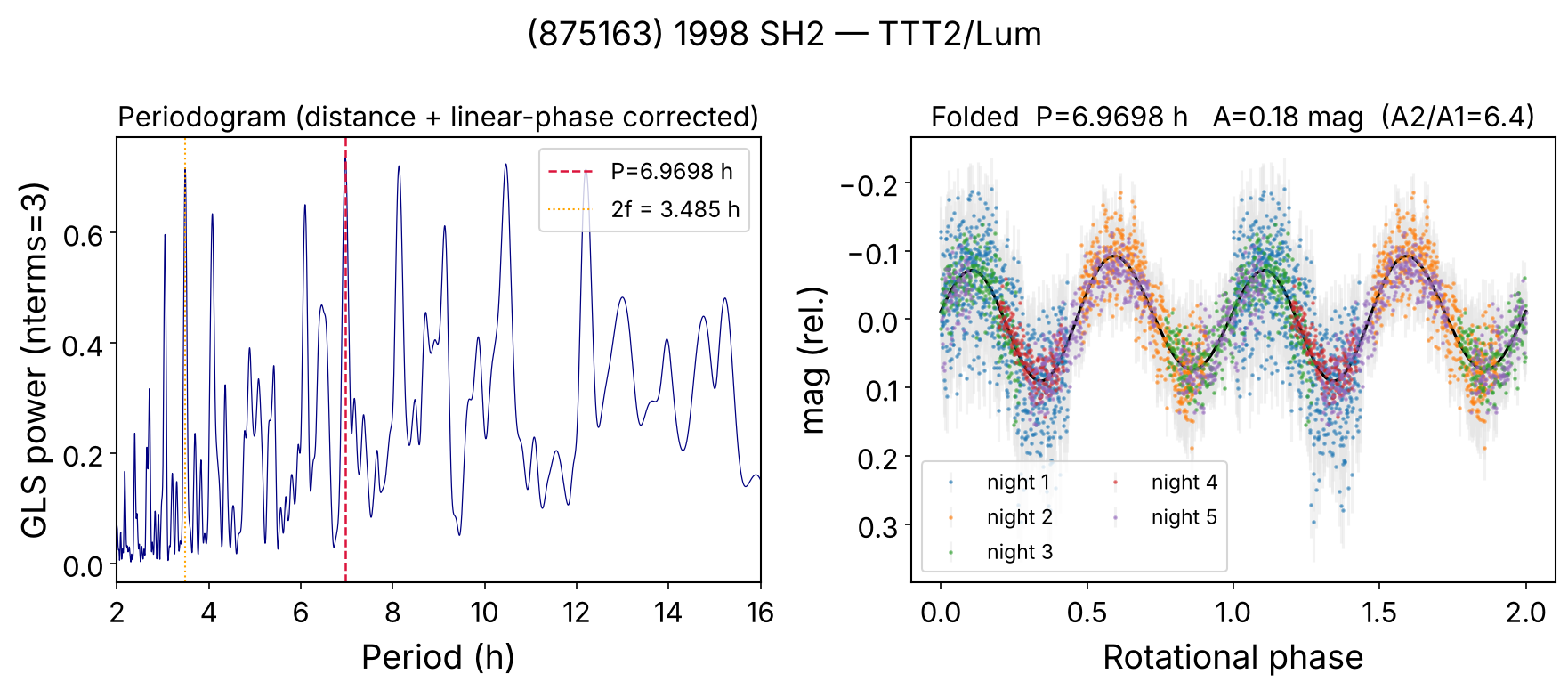}
\caption{Final GLS periodogram (left) of the distance- and
linear-phase-corrected magnitudes, and the light curve folded at
$P_{\rm rot} = 6.970$\,h (right), coloured by night with the three-harmonic
model overplotted. The dominant second harmonic produces the bimodal,
two-maxima morphology.}
\label{fig:lc}
\end{figure}

The light curve is strongly bimodal, the second harmonic dominating the first
($A_2/A_1 \simeq 6.4$), with a peak-to-peak amplitude of $\simeq 0.18$\,mag
(Fig.~\ref{fig:lc}). 

Throughout, the Fourier series was truncated at three harmonics because it is the number that
minimizes the Bayesian information criterion; this choice does not affect our
results, as the second-to-first amplitude ratio is stable against it
($A_2/A_1 = 6.3$--$6.4$ for two to five harmonics) and the third and higher
harmonics are not significant ($A_3 \simeq 0.006$\,mag, $A_4, A_5 \lesssim
0.002$\,mag). The bimodality is not an artefact of the noisier first night:
repeating the whole period search with night~1 excluded leaves the period
unchanged ($6.972$\,h) and the amplitude ratio essentially so
($A_2/A_1 = 6.1$, peak-to-peak $0.177$\,mag versus $0.183$\,mag), so the
two-maxima morphology is a robust property of the light curve.

\subsection{Phase behaviour}\label{sec:phase}
\begin{figure}
\centering
\includegraphics[width=\columnwidth]{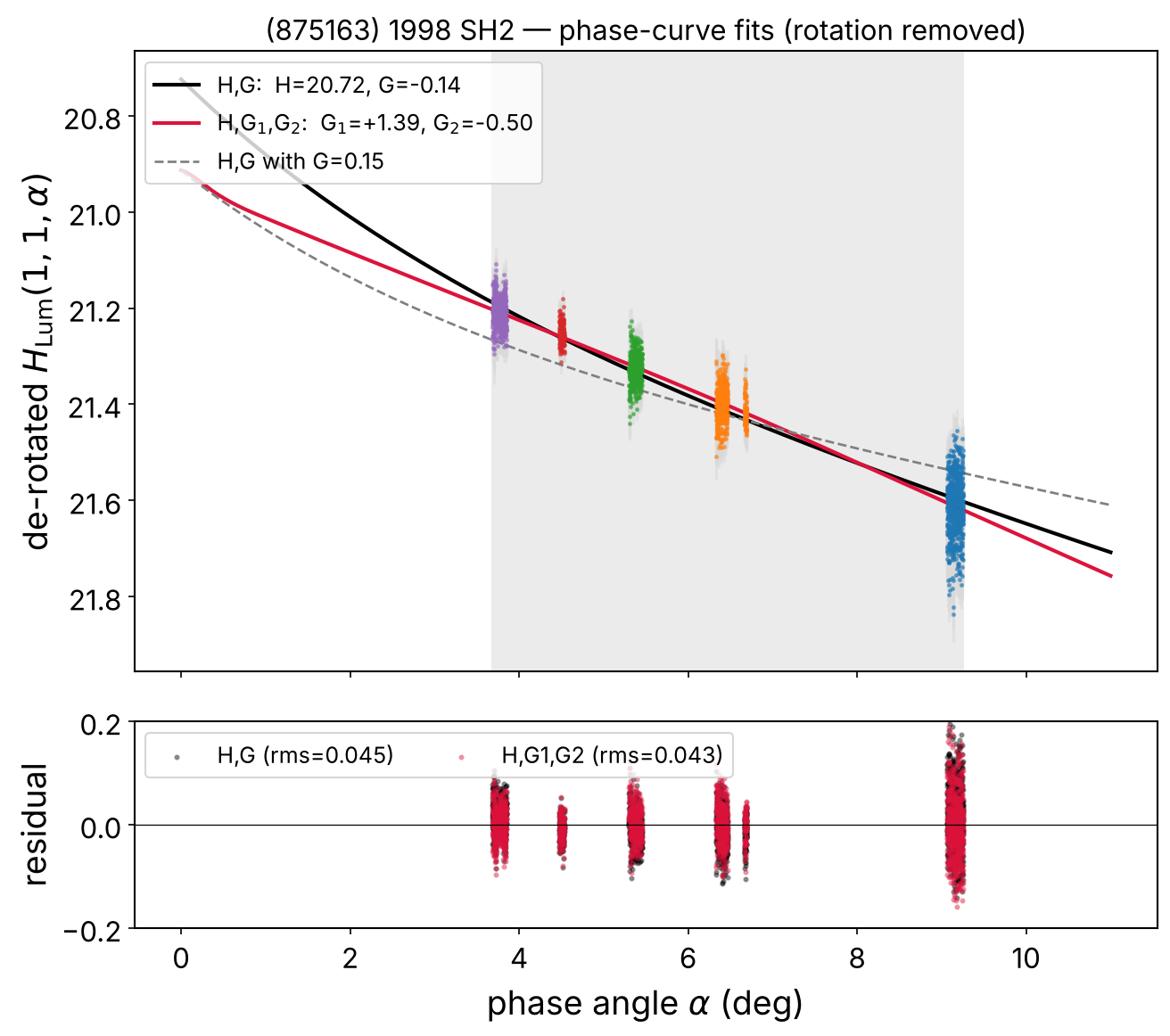}
\caption{De-rotated phase curve with the $H,G$ and $H,G_1,G_2$ fits (top) and
their residuals (bottom). Within the observed range (shaded) the models are
indistinguishable; they diverge only below $\alpha \simeq 3.7\degr$, where there
are no data. The recovered slope parameters are unphysical, showing the fit is
unconstrained. The plotted magnitudes are de-rotated: the fitted three-harmonic
rotation has been removed, which reduces the within-night r.m.s.\ by roughly a
factor of two (e.g.\ from $0.080$ to $0.037$\,mag on night~2). The vertical
spread that remains is dominated by the phase-angle trend itself (the quantity
being fitted) together with the $\simeq 0.04$\,mag residual scatter, and not by
the rotational amplitude, which has been subtracted.}
\label{fig:phasefit}
\end{figure}

Knowing the period, we separated rotation from the phase trend by a joint linear
fit of a three-harmonic Fourier series (at $P_{\rm rot}$) plus a smooth term in
$\alpha$, and removed the periodic part to obtain a de-rotated $H_{\rm Lum}(1,\alpha)$.
Fitting the IAU $H,G$ and $H,G_1,G_2$ systems \citep{bowell1989,muinonen2010}
with the \textsc{sbpy} implementation \citep{sbpy} returns \emph{unphysical}
slope parameters ($G = -0.14$ with $\mathrm{corr}(H,G) = +0.96$; $G_1 = +1.39$,
$G_2 = -0.50$ with $\mathrm{corr}(G_1,G_2) = -0.99$), the signature of a fit
that the data do not constrain (Fig.~\ref{fig:phasefit}). The two models agree
to 0.015\,mag within the observed range and diverge only below
$\alpha \simeq 3.7\degr$, where there are no data: with the opposition surge
($\alpha \lesssim 2\degr$) unsampled, the slope parameter has no leverage. In
the $H,G$ system the data correspond to $G \lesssim 0$; this formally negative
value merely reflects a phase curve steeper than the $G=0.15$ default and should
not be assigned physical significance. The one-parameter form is a poor
descriptor over so narrow
a range, so we report a linear coefficient instead.

\begin{figure}
\centering
\includegraphics[width=\textwidth]{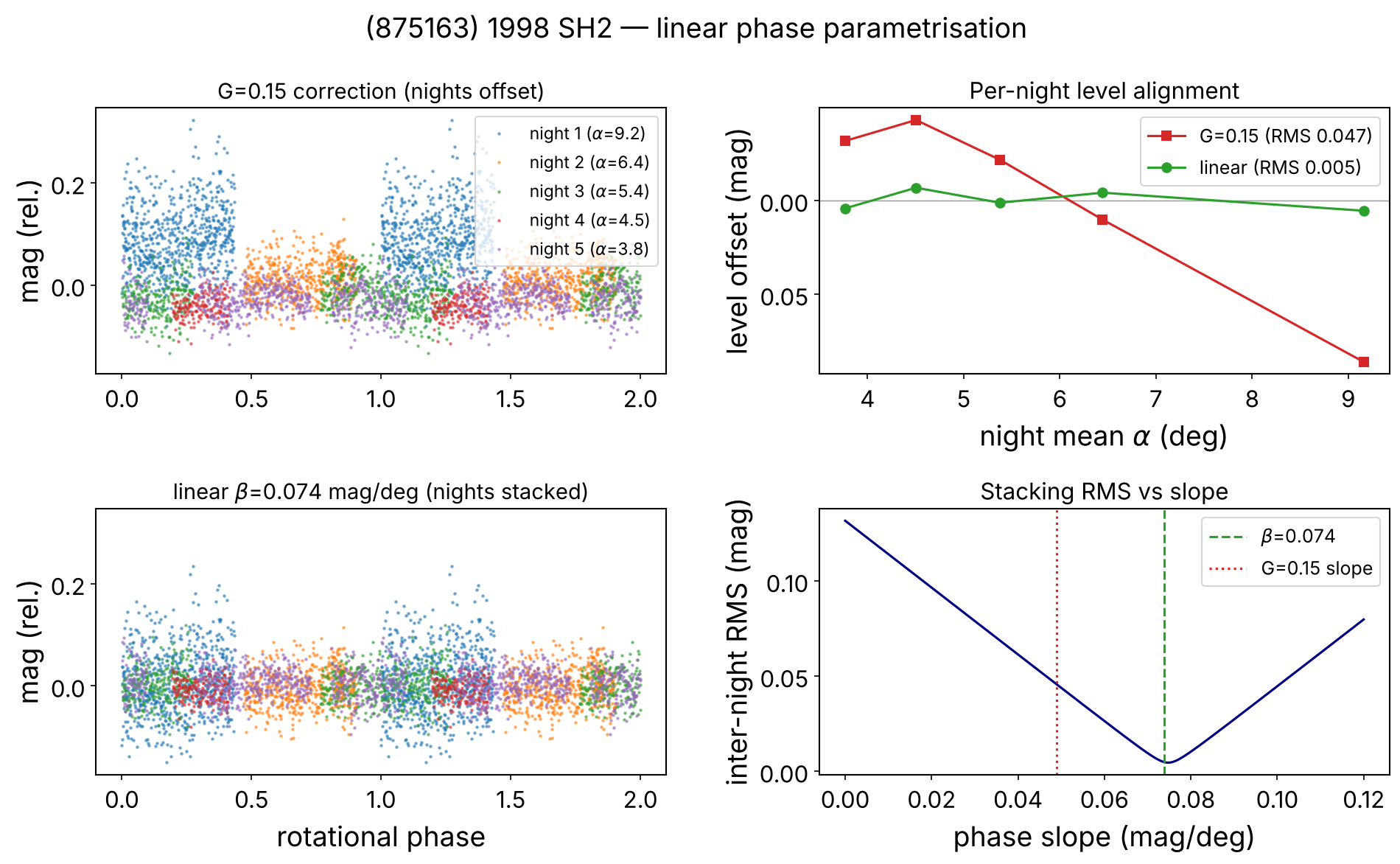}
\caption{Linear phase parametrization. Left: the light curve folded with the
$G = 0.15$ correction (top, nights offset) versus the linear $\beta_{\rm app}$ correction
(bottom, nights stacked). Right: per-night level offset versus phase angle, and
the inter-night r.m.s.\ as a function of assumed phase slope, minimized at
$\beta_{\rm app} \approx 0.074$\,mag\,deg$^{-1}$. This curve reflects the internal
inter-night alignment only and does not include absolute night-to-night
calibration systematics; the gradient is treated as an upper limit to the
intrinsic slope owing to the object's weak, rising activity.}
\label{fig:linphase}
\end{figure}

Over this essentially linear arc the measurable descriptor is a single phase
coefficient. Fitting $H_{\rm Lum}(1,1,\alpha) = H_{\rm Lum}(1,1,0) + \beta_{\rm app}\,\alpha$ jointly
with the rotation gives an apparent inter-night gradient
$\beta_{\rm app} \approx 0.074$\,mag\,deg$^{-1}$ (jackknife range over the five nights
$0.070$--$0.076$), the value that co-registers the nights: after de-rotation the
per-night level offsets fall from an r.m.s.\ of 0.047\,mag with the $G = 0.15$
correction (a clear trend with $\alpha$, reaching $+0.086$\,mag on the
highest-$\alpha$ night) to 0.005\,mag with the linear correction
(Fig.~\ref{fig:linphase}). The formal uncertainty on $\beta_{\rm app}$ from propagating the per-point photometric
errors, $\sim\!0.0004$\,mag\,deg$^{-1}$, is unrealistically small: as an
inter-night gradient, $\beta_{\rm app}$ is limited not by the point-to-point scatter but by
the night-to-night stability of the photometric zero-point. Because the object
moves between fields, each nightly level carries a common zero-point systematic
$\sigma_{\rm ZP}$ that averaging within a night does not reduce; over the five
nights, which span a phase-angle lever arm
$L = [\sum_n(\alpha_n-\bar\alpha)^2]^{1/2} = 4.2\degr$, this propagates into a
slope uncertainty $\sigma_{\beta_{\rm app}} \simeq \sigma_{\rm ZP}/L$. A conservative
$\sigma_{\rm ZP} \approx 0.02$\,mag then gives
$\sigma_{\beta_{\rm app}} \approx 0.005$\,mag\,deg$^{-1}$, an order of magnitude above the
formal value and the dominant term in the error budget. We therefore quote
$\beta_{\rm app} = 0.074 \pm 0.005$\,mag\,deg$^{-1}$ for the apparent inter-night gradient; because
rising coma can only steepen it, this remains an upper limit to the intrinsic
nuclear phase slope, $\beta_{\rm nuc}$.

This apparent gradient is steep. It exceeds the median linear coefficient of JFC nuclei,
$\beta_{\rm nuc} = 0.046$\,mag\,deg$^{-1}$ \citep{kokotanekova2017}, and the mean
$0.053 \pm 0.016$\,mag\,deg$^{-1}$ of \citet{snodgrass2011}, lying toward the upper end of the observed range
(individual JFC nuclei span $\sim 0.025$--$0.096$\,mag\,deg$^{-1}$;
\citealt{snodgrass2011,kokotanekova2017}). Three systematics suggest that it should be regarded as an upper limit to the intrinsic surface slope:
\begin{enumerate}
\item \emph{Residual coma.} 1998\,SH$_2$ was weakly and increasingly active
  during the run \citep{farnocchia2026}. Because our later nights are both at
  smaller $\alpha$ and later in time (rising coma), a coma that brightens with
  time steepens the apparent inter-night gradient. Quantitatively, a coma
  brightening of $\Delta C$ mag over the run would bias the gradient by
  $\simeq \Delta C/\Delta\alpha$, with $\Delta\alpha \simeq 5.6\degr$; for
  $\Delta C = 0.02$, $0.05$ and $0.10$\,mag this is $0.004$, $0.009$ and
  $0.018$\,mag\,deg$^{-1}$. A change of only $\sim\!0.05$--$0.10$\,mag --
  plausible given the growth from $0.06$ to $0.24$\,mag between 13 and 30
  September -- would thus account for a substantial fraction of the measured
  $0.074$\,mag\,deg$^{-1}$. Tellingly, one of the steepest JFC
  nucleus coefficients reported, $\beta_{\rm nuc} = 0.096 \pm 0.004$ for 47P
  \citep{kokotanekova2017}, was itself derived from data taken while the comet
  was probably active. We stress, however, that this is only a cautionary
  analogy and not evidence that a steep slope \emph{implies} activity: 47P was
  observed over several apparitions and its activity need not have been changing
  within any single run, whereas in our data the coma demonstrably grows along
  the arc. The robust point is simply that an increasing coma can steepen an
  \emph{apparent} slope, as it may here. 
  We searched for this excess directly in deep stacks of the first and last
nights (Section~\ref{sec:results}, Fig.~\ref{fig:stacks}).
\item \emph{Rotational modulation.} With partial rotational coverage on the
  short nights, light-curve amplitude can leak into $\beta_{\rm app}$. We mitigated this
  by fixing the period and removing the rotational modulation before (and
  jointly with) the phase fit -- the precaution that steepened the 162P
  coefficient from $0.047$ to $0.051$ once a shape model was applied
  \citep{donaldson2023}.
\item \emph{Phase interval.} Our arc begins at $3.7\degr$, so the intercept is
  an extrapolation; we therefore report it as $H_{\rm Lum}(1,1,0)$ over the stated
  range rather than as a definitive absolute magnitude.
\end{enumerate}

Crucially for that extrapolation, available JFC nucleus phase curves show
little or no detectable opposition surge and are well represented by linear phase functions down to sub-degree
angles (162P shows none to $0.39\degr$; \citealt{snodgrass2011,kokotanekova2017,
donaldson2023}), the opposite of the asteroidal trend in which darker surfaces
show weaker but still non-zero surges \citep{belskaya2000}. Adopting this
population prior, the linear law can be carried to opposition, giving
$H_{\rm Lum}(1,1,0) \approx 20.9$--$21.0$ over $3.7\degr$--$9.3\degr$ in the
Luminance band -- an estimate otherwise unavailable, but still carrying the
Luminance zero point and a small (weak-coma) contribution.

Applying the linear correction and re-running the periodogram homogenizes the
five nights and cleans the spectrum: 6.970\,h becomes the \emph{global} GLS
maximum rather than a secondary peak, and $A_2/A_1$ rises from 3.6 to 6.4 as the
spurious inter-night trend that had leaked into the odd harmonic is removed. 
This procedure is self-consistent: re-deriving the period from the
phase-corrected magnitudes and re-fitting $\beta_{\rm app}$ converges in a single
iteration, with a period shift of only $\sim$0.003\,h -- more than an order of
magnitude smaller than $\sigma_P$ -- and $\beta_{\rm app}$ unchanged.

\section{Results and discussion}
\label{sec:results}
The single-apparition solution for (875163)\,1998\,SH$_2$ is: a synodic rotation
period $P_{\rm rot} = 6.970 \pm 0.020$\,h  (to our knowledge the first
reported for this object) a bimodal light curve ($A_2/A_1 \simeq 6.4$) of
peak-to-peak amplitude $\simeq 0.18$\,mag, and an apparent inter-night gradient
$\beta_{\rm app} = 0.074 \pm 0.005$\,mag\,deg$^{-1}$, an upper limit to the
nuclear slope ($\beta_{\rm nuc} \lesssim 0.074$). The light-curve amplitude sets a
lower limit on the elongation, $a/b \ge 10^{0.4\,\Delta m} \simeq 1.18$ for the
ratio of the projected axes, assuming a triaxial body whose brightness follows
its cross-section and an equatorial aspect; a non-equatorial view would imply a
larger true elongation, so this bound is conservative. Any coma present during
the run further dilutes the rotational modulation, lowering the observed
amplitude relative to that of the bare nucleus; the inferred elongation is
therefore a lower limit on this count as well, and a quiescent apparition could
reveal a larger amplitude and a more elongated shape.

The period is consistent with, and refines, the radar constraints. The Goldstone
echo bandwidth is $B = 4\upi D_{\rm eff}\cos\delta / (\lambda P)$, where
$D_{\rm eff}$ is the pole-on breadth, $\delta$ the subradar latitude and
$\lambda = 3.5$\,cm. A spherical, equatorial 380\,m body ($D_{\rm eff} = D$,
$\cos\delta = 1$) with $B = 7$\,Hz gives $P = 5.4$\,h, the reported upper bound.
Substituting our measured period instead yields
$D_{\rm eff}\cos\delta \approx 490$\,m. Since $\cos\delta \le 1$, this is a lower
bound, $D_{\rm eff} \gtrsim 490$\,m, on the pole-on breadth -- larger than the
NEOWISE mean diameter ($380 \pm 57$\,m) and therefore indicating a non-spherical,
elongated body. We stress that this is a geometric consistency argument rather
than a shape measurement: the subradar latitude $\delta$ is unknown, the thermal
diameter carries a $\sim\!15$ per cent uncertainty, and the NEOWISE
area-equivalent diameter and the radar projected breadth are not the same
quantity, so their ratio does not translate directly into an axis ratio. For an
equatorial view and the nominal diameter it would correspond to a characteristic
breadth $\sim\!1.3$ times the thermal-equivalent diameter, but the robust
statement is simply that the radar breadth ($\gtrsim 490$\,m) and the light-curve
amplitude ($a/b \gtrsim 1.18$) are independently and mutually consistent with an
elongated nucleus. This removes the apparent tension with the 5.4\,h
``upper bound'', which assumed sphericity. The reported absence of
significant radar-bandwidth variation over 1.06\,h is likewise consistent with a
6.97\,h period.

The period also bears on the origin of the activity. A strengthless, gravity-dominated
body sheds mass from its equator when it spins faster than the critical period
$P_{\rm crit} = \sqrt{3\pi/(G\rho)}$, where $\rho$ is the bulk density; equivalently,
a given period becomes critical only below a density
$\rho_{\rm crit} = 3\pi/(G P_{\rm rot}^2)$. For $P_{\rm rot} = 6.970$\,h this is
$\rho_{\rm crit} \approx 0.22$\,g\,cm$^{-3}$, lower than the bulk densities measured for
cometary nuclei (e.g.\ $0.5$\,g\,cm$^{-3}$ for 67P; \citealt{kokotanekova2017} and
references therein). At a representative nucleus density of $0.3$--$0.6$\,g\,cm$^{-3}$ the
critical period is $6.0$--$4.7$\,h, comfortably shorter than $6.970$\,h, and modest
elongation ($a/b \gtrsim 1.18$) lengthens $P_{\rm crit}$ further at the long-axis tips.
The spin is therefore well below the rotational-disruption limit for any plausible
density: rotation is too slow to drive the observed mass loss centrifugally, so the
weak activity of 1998\,SH$_2$ must be sustained by sublimation rather than by rotational
instability. This complements the dynamical \citep{farnocchia2026} and taxonomic evidence
for a genuinely volatile-bearing, comet-like body.

\begin{figure}
\centering
\includegraphics[width=\textwidth]{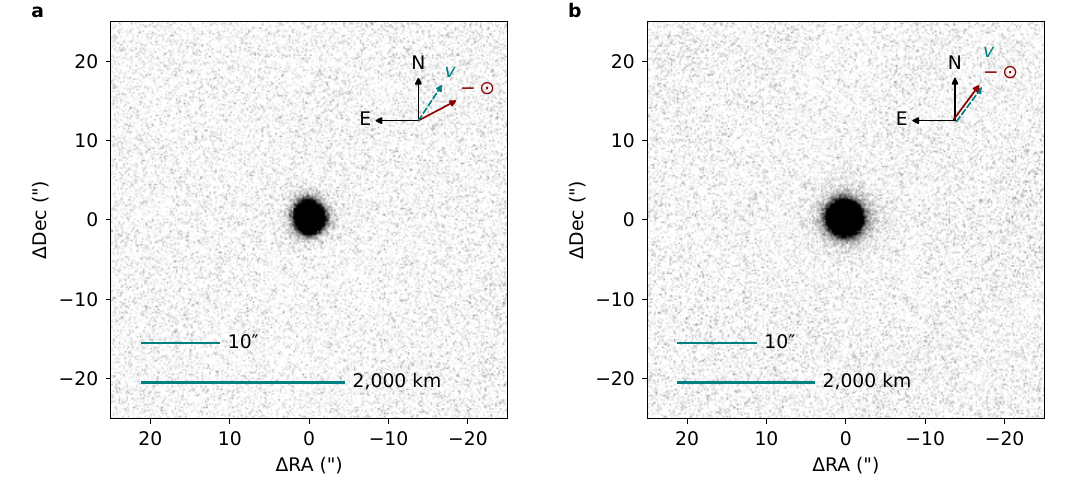}
\caption{Co-moving stacks of 1998~SH$_2$ obtained with TTT2 in the Luminance filter on
(a) 2025 September 9 ($N = 698$ frames, 175~min total exposure) and (b) 2025 September 14 ($N = 509$ frames,
255~min total exposure).
Each frame was aligned on the object (ephemeris plus a smooth recentring fit;
Section~\ref{sec:results}) and combined with a weighted, $\sigma$-clipped mean. North is up and east
to the left. Arrows indicate the antisolar direction and the
apparent sky motion of the object. Scale bars show $10''$ and 2000~km at the geocentric distance of each
night.}
\label{fig:stacks}
\end{figure}

To search for extended emission, we built one deep stack for each of the first and last
nights of the run (Fig.~\ref{fig:stacks}), from individual exposures of 15~s and 30~s,
respectively. Frames were selected by photometric transparency (zero-point within
0.3~mag of the nightly 90th percentile) and seeing (FWHM below the nightly median plus
three times its median absolute deviation). In each frame a sky plane was fitted and
subtracted with the object masked, and field sources were masked to suppress residual
star trails. The object was then located by a windowed centroid around its JPL Horizons
position. The centroid--ephemeris offsets were modelled as a linear function of time
with iterative $3\sigma$ rejection, so that frames in which the object was blended with
a field star do not bias the alignment; the r.m.s. of the retained centroids about the
model is $0.12''$ and $0.07''$. Each frame was resampled with a single affine
transformation onto a north-up grid centred on the modelled position, scaled to a
common photometric zero-point and combined with a weighted $\sigma$-clipped mean.

\begin{figure}
\centering
\includegraphics[width=\textwidth]{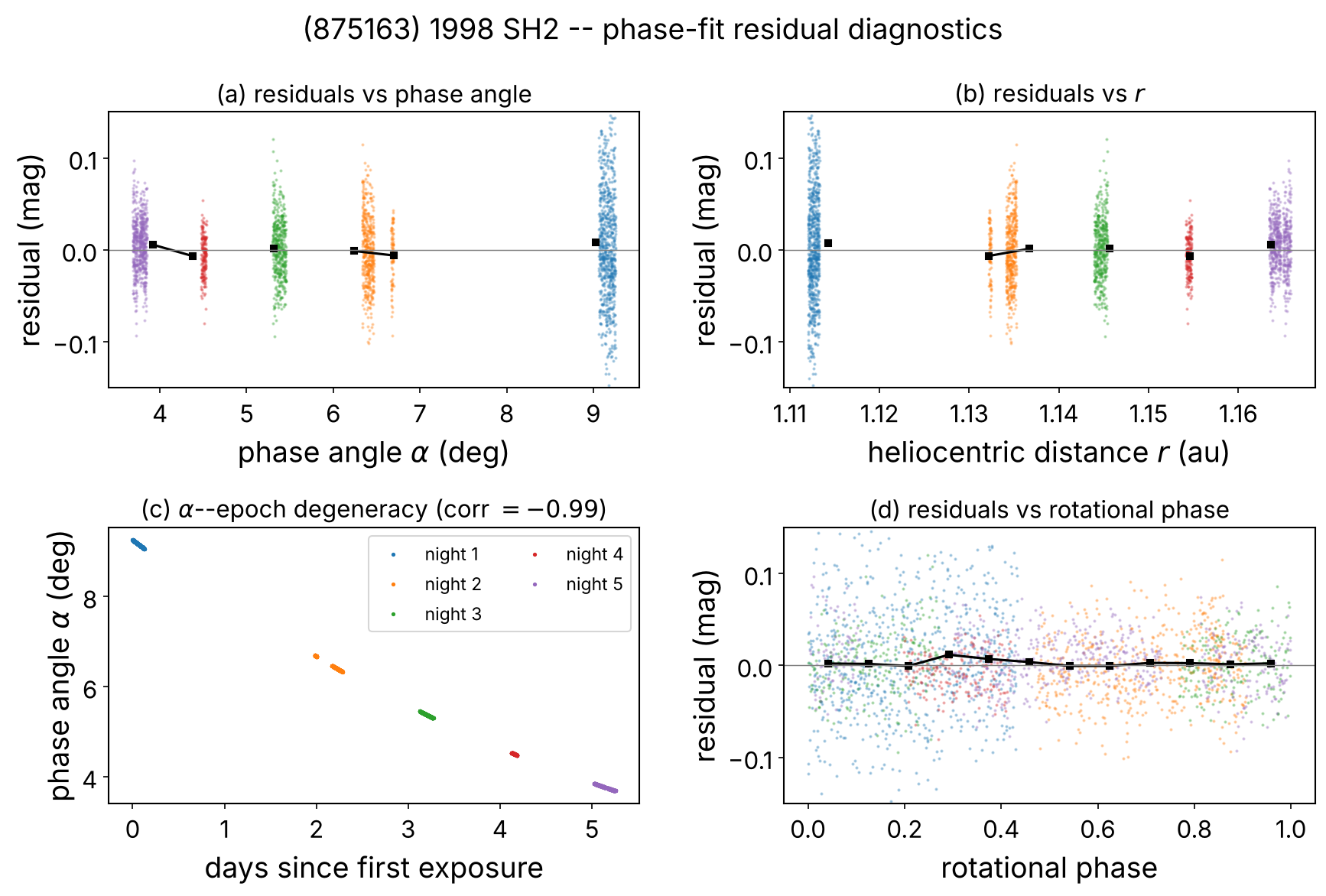}
\caption{Residual diagnostics of the joint rotation + linear-phase fit
(r.m.s.\ 0.043\,mag), coloured by night. (a) Residuals versus phase angle, (b)
versus heliocentric distance (a proxy for activity level / true anomaly) and (d)
versus rotational phase are all flat; in these three panels the black squares
are the mean residuals in twelve equal bins of the abscissa (plotted where a bin
contains more than three points), and show no obvious remaining trends over the
sampled parameter range. (c) Phase angle versus epoch: the near-perfect anti-correlation
(corr $=-0.99$) is the degeneracy that prevents separating an intrinsic phase
slope from a time-dependent coma brightening with these data alone.}
\label{fig:diag}
\end{figure}

No tail or coma is detected in either stack. 
To check that the object is not resolved, we measured its FWHM in the
co-moving stacks and found it consistent, within the uncertainties, with the
mean stellar FWHM of each night (the seeing listed in Table~\ref{tab:obs}); the
object is therefore consistent with a point source on both nights.
This non-detection is
consistent with the weak activity reported by \citet{farnocchia2026}, whose
tail detection on September 13 required larger apertures and who estimate a
coma contribution of only $\sim$0.06~mag at that date; a coma at that level,
concentrated within the seeing disc, lies below the sensitivity of our
profiles. Crucially for the increasing-activity hypothesis of
Section~\ref{sec:phase}, the object's FWHM is unchanged between the first and last night and no
brightness excess is seen in either stack: within our sensitivity the coma
neither appears nor grows across the run. The deep
stacks therefore cannot by themselves confirm or rule out the mild
brightening inferred from the phase analysis, consistent with treating
$\beta_{\rm app}$ as an upper limit to the nuclear slope.
The steep phase gradient is in the sense expected for a dark cometary surface and
is close to values reported for individual JFC nuclei. Fig.~\ref{fig:diag} shows
the relevant diagnostics for our data: the fit residuals are flat against phase
angle, heliocentric distance and rotational phase (panels a, b, d; r.m.s.\
0.043\,mag), showing no obvious remaining trends over the sampled range.
Panel (c), however, makes explicit the near-perfect anti-correlation between
phase angle and epoch (corr $=-0.99$), which leaves an intrinsic phase slope and
a time-dependent (coma) brightening inseparable from these data alone. Crucially,
this ambiguity does not materially affect the rotation-period determination or
the light-curve amplitude: $\beta_{\rm app}$ enters the analysis only to
co-register the nightly zero-points and is not interpreted as a surface property,
so the period, the amplitude and the resulting shape constraint are independent
of the phase slope. It does affect $\beta_{\rm app}$ and the extrapolated
$H_{\rm Lum}$ themselves, which is precisely why we treat both as bounded rather
than measured quantities. We therefore quote $\beta_{\rm app} = 0.074 \pm 0.005$\,mag\,deg$^{-1}$ as an
empirical co-registration coefficient and an upper limit to the nuclear slope,
$\beta_{\rm nuc} \lesssim 0.074$ (steep slopes derived from active-comet data are known
to be revised when only inactive data are used) and leave a physical phase-slope
determination to a future, demonstrably inactive apparition, when the surface can
be characterized free of coma.

\section{Conclusions}
\label{sec:conclusions}
Dense TTT2 photometry taken during the 2025 apparition of the near-Earth comet
(875163)\,1998\,SH$_2$ yields its first rotation period,
$6.970 \pm 0.020$\,h, and a bimodal light curve of $\simeq 0.18$\,mag
amplitude. Combined with the published radar bandwidth and NEOWISE diameter, the
period implies a projected radar breadth $\gtrsim 490$\,m, larger than the mean
thermal diameter and thus indicating an elongated nucleus, consistent with the
light-curve amplitude and resolving the radar's period ambiguity.

Two methodological points generalize. First, with only five nights the
periodogram is alias-dominated, and the true period is secured not by peak
height but by demanding coherence on the continuous nights. Secondly, a single
apparition confined to $\alpha = 3.7\degr$--$9.3\degr$ does not constrain the
$H,G$ or $H,G_1,G_2$ slope parameters, which run to unphysical values with
near-perfect degeneracy; the default $G = 0.15$ measurably fails to co-register
the nights. A linear phase coefficient is the appropriate description, but for
1998\,SH$_2$ its value ($\beta_{\rm app} = 0.074 \pm 0.005$\,mag\,deg$^{-1}$) -- steep, in the
upper part of the JFC distribution -- is best read as an upper limit, because
the object was weakly and increasingly active during the run. Under the
population prior of little or no opposition surge in available JFC nucleus phase
curves, the linear law
extrapolates to $H_{\rm Lum}(1,1,0) \approx 20.9$--$21.0$ (Luminance) over the
observed range. A physical slope, a colour-corrected $H$, and a clean separation
of surface and coma contributions will require photometry to low phase angle
and, ideally, a quiescent apparition.

\section*{Acknowledgements}
This work is based on observations made with the Two-meter Twin Telescope (TTT), located at the Teide Observatory of the Instituto de Astrofísica de Canarias (IAC) and operated by Light Bridges in Tenerife, Canary Islands, Spain. The observation time rights (DTO) used for this research were consumed in the PEI "SIDERA2025". This research used storage and computing capacity in ASTRO POC's EDGE computing center at Tenerife under the form of Indefeasible Computer Rights (ICR), consumed in the PEI "SIDERA2025". Dr. Antonio Maudes’s insights in economics and law were instrumental in shaping the development of this work. Data reduction used the \textsc{GPUPhot} pipeline  \citep{Alarcon2026GPUPhot, LemesPerera2026GPUPhot}.
We acknowledge financial support from the Agencia Estatal de Investigaci\'on del Ministerio de Ciencia e Innovaci\'on (AEI-MCINN) under the grant "Hydrated Minerals and Organic Compounds in Primitive Asteroids 2" with reference PID2024-160618NB-C22 and through the Severo Ochoa Centre of Excellence accreditation awarded to the Instituto de Astrofísica de Canarias, grant CEX2025-001609-S, funded by MICIU/AEI/10.13039/501100011033.

\section*{Data Availability}
The full table of 2392 calibrated photometric measurements underlying this
article (light-time-corrected epoch, Luminance magnitude and uncertainty, and
the $r$, $\Delta$ and $\alpha$ of each exposure) is provided as Supplementary
Material with the online version of the article and is also deposited in the
Zenodo public repository (\url{https://doi.org/10.5281/zenodo.23978129}).

\section*{Declaration of generative AI and AI-assisted technologies in the manuscript preparation process}
During the preparation of this work the authors used Claude (Anthropic) to assist
with the development and checking of the data-analysis and figure-generation
code, and to improve the readability and language of the manuscript. After using
this tool, the authors reviewed and edited the content as needed and take full
responsibility for the content of the publication.

\appendix

\section{Nightly light curves}\label{app:nightly}
\begin{figure}
\centering
\includegraphics[width=\textwidth]{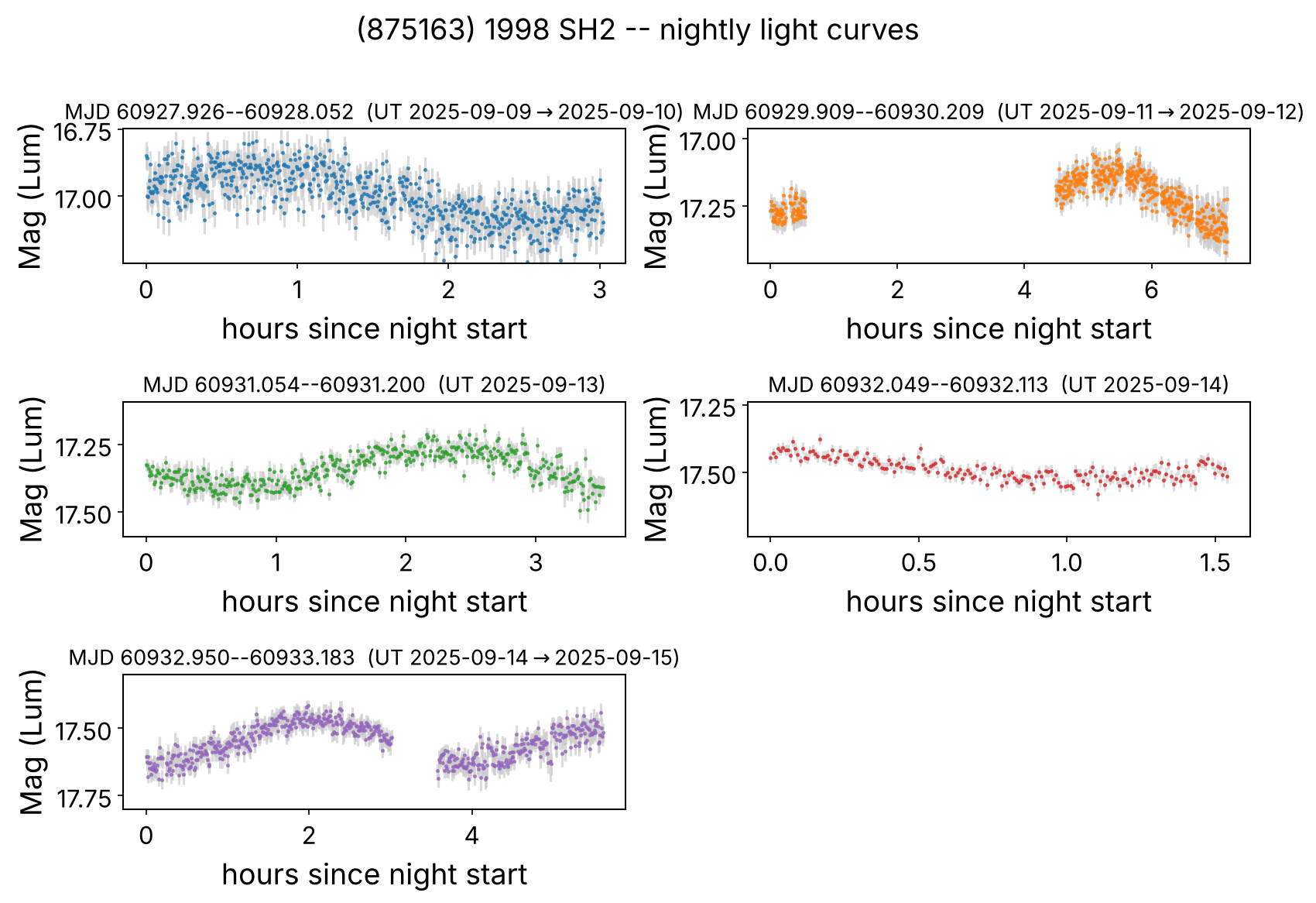}
\caption{Light curves of (875163)\,1998\,SH$_2$ obtained on each observing night:
observed magnitude (Luminance filter) versus time since the night's first
exposure, each panel centred on the night's mean magnitude and spanning
0.5\,mag. Panel titles give the MJD range and the spanned UT date(s); an arrow
denotes a midnight (integer-MJD) crossing.}
\label{fig:nightly}
\end{figure}

Fig.~\ref{fig:nightly} shows the light curve obtained on each of the five nights,
plotted against time elapsed since the first exposure of that night. This
elapsed-time axis stays continuous even when a night crosses midnight (an
integer-MJD boundary): the title of each panel gives the true MJD range and the
UT date(s) spanned, and nights 1, 2 and 5 each cross such a boundary.

\end{document}